\documentclass[prl,twocolumn,noshowpacs,preprintnumbers,amsmath,amssymb]{revtex4-1}

\usepackage{amsfonts,amsthm,amsmath,amssymb,mathrsfs,mathtools}
\usepackage{dsfont}
\usepackage{dcolumn}
\usepackage{epsfig, color} 
\usepackage{verbatim,hyperref,graphics,graphicx,textcomp}
\usepackage{epstopdf}
\usepackage{xr}
\usepackage{enumitem}
\usepackage[para,flushleft]{threeparttable}

\begin{document}

\title{Model of inverse bleb growth explains giant vacuole dynamics during cell mechanoadaptation}

\author{Andrea Cairoli}\email{andrea.cairoli@crick.ac.uk}
\altaffiliation{Current author’s affiliation: The Francis Crick Institute, 1 Midland Road, London NW1 1AT, United Kingdom}
\author{Alice Spenlehauer}
\author{Darryl R Overby}
\author{Chiu Fan Lee}\email{c.lee@imperial.ac.uk}
\affiliation{Department of Bioengineering, Imperial College London, London SW7 2AZ, UK}

\begin{abstract}

Cells can withstand hostile environmental conditions manifest as large mechanical forces such as pressure gradients and/or shear stresses by dynamically changing their shape. Such conditions are realized in the Schlemm's canal of the eye where endothelial cells that cover the inner vessel wall are subjected to the hydrodynamic pressure gradients exerted by the aqueous humor outflow. These cells form fluid-filled dynamic outpouchings of their basal membrane called \textit{giant vacuoles}. The inverse of giant vacuoles are reminiscent of cellular blebs, extracellular cytoplasmic protrusions triggered by local temporary disruption of the contractile actomyosin cortex. Inverse blebbing has been first observed experimentally during sprouting angiogenesis, but its underlying physical mechanisms are poorly understood. Here, we identify giant vacuole formation as inverse blebbing and formulate a biophysical model of this process. Our model elucidates how cell membrane mechanical properties affect the morphology and dynamics of giant vacuoles and predicts coarsening akin to Ostwald ripening between multiple invaginating vacuoles. Our results are in qualitative agreement with observations from the formation of giant vacuoles during perfusion experiments. Our model not only elucidates the biophysical mechanisms driving inverse blebbing and giant vacuole dynamics, but also identifies universal features of the cellular response to pressure loads that are relevant to many experimental contexts. 

\end{abstract}

\maketitle

\section{Significance statement}

Human Schlemm's canal endothelial cells in physiological conditions are subjected to a pressure gradient caused by the flow of aqueous humor in the basal-to-apical direction across the endothelium leading to the formation of cellular outpouchings called giant vacuoles. The physical mechanisms regulating giant vacuole formation are unknown. By describing giant vacuoles as inward blebs, we formulate a model of their growth and collapse that captures the characteristic features observed experimentally. Our theory reveals that the abrupt increase in surface tension caused by membrane stretching, which is required to accommodate the large areal strains locally induced by inward blebbing, limits giant vacuole growth. The model also predicts a competition between multiple invaginating vacuoles in which big vacuoles win over small vacuoles.

\section{Introduction}

Morphological regulation is crucial for the survival of eukaryotic cells in physiological conditions \cite{Alberts2002}. Shape changes are involved in a broad spectrum of cellular functions, such as motility and division \cite{Boal2012,Huang2012}, and represent the main cellular response to large mechanical or osmotic pressure gradients and shear stresses \cite{Vogel2006}. Exemplary forces of this type are found especially in the Schlemm's canal (SC) of the human eye. SC belongs to the conventional outflow pathway of aqueous humor \cite{Johnson2000}, a transparent fluid that bathes the lens and other tissues of the eye and is the major determinant of intraocular pressure. As elevated intraocular pressure is a risk factor for glaucoma \cite{Grant1951}, its regulation is a key physiological process, primarily realized by draining aqueous humor through the inner wall endothelium of SC \cite{Gong1996}. This is a continuous monolayer of spindle-shaped endothelial cells (ECs), that are attached between one another by tight junctions and adhere to a discontinuous basement membrane \cite{Bill1972}. These cells are subjected to the hydrodynamic pressure exerted by the aqueous humor outflow (typically about 1 mmHg in healthy human eyes \cite{Ethier2004}), which corresponds to a significant force (in cellular terms) in the basal-to-apical direction (about 60 nN \cite{Overby2011}).   
 
Under these conditions, SC ECs form large dome-like fluid-filled outpouchings, called giant vacuoles (GVs) \cite{Holmberg1959,Garron1958,Speakman1959}, by locally invaginating the basal aspect of the cell surface into the cellular body. These structures are typically oriented along the cell axis and are entirely bounded by a smooth unit membrane, i.e., their inner cavity remains extracellular at all times. They thus exhibit a "signet ring" shape: The cell appears as a thin, continuous lining around the GV cavity with the nucleus bulging to one side. Interestingly, these structures are not unique to SC ECs but can also be found in other cell types \cite{Pedrigi2011,Tripathi1977,Tripathi1973,Grzybowski2006}. 

GVs have been studied experimentally both in vivo and in vitro (see reviews \cite{Overby2011,Ethier2002}). All these experiments revealed several characteristic features of GVs, here briefly summarized. 
\begin{enumerate}[label=(\roman*)]
\item{The formation of GVs depends on the pressure drop $\Delta p$ across the inner wall endothelium that drives the growth process of GVs \cite{Tripathi1971,Inomata1972,Tripathi1974,Grierson1977}. When a positive $\Delta p$ is established, GVs formed within circa 15 min \cite{Ainsworth1990}; upon removal of the pressure drop, they shrink within a timescale of the order of minutes \cite{vanBuskirk1974,Brilakis2001} (survival time).}
\item{GVs typically induce large increase of the net surface area of SC cells (by threefold to fourfold per vacuole when the intraocular pressure is increased from 15 mmHg to 30 mmHg according to the data in \cite{Grierson1977}; see details in \cite{Overby2011}).} 
\item{Their morphology depends on the cell surface tension: reduction of the cortical tension through impairment of cellular actomyosin contractility leads to larger GVs \cite{Sabanay2000}.}
\end{enumerate}

Despite the abundance of experimental studies, however, the biophysical mechanisms underlying GV formation have not yet been fully understood. 
To accommodate the large shape deformations associated with GVs, cells regulate their surface area through a complex dynamical process, which involves remodeling of the cell membrane and reorganization of the actomyosin cortex, orchestrated by the surface tension \cite{Morris2001,Keren2008,Batchelder2011,Boulant2011,Houk2012,Diz2013}. 
This process promotes the formation of many different cellular protrusions, such as blebs \cite{Cunningham1995,Charras2008,Tinevez2009}, 
microvescicles \cite{Sens2007} and either tubular or spherical membrane invaginations (reminiscent of GVs) \cite{Morris2001,Morris2003,Staykova2013}. 
For the underlying dynamical picture, a natural timescale of the order of minutes has been reported \cite{kosmalska2015}: 
At shorter times the protrusions are rapidly formed via a local passive mechanical process; on longer times they are reabsorbed via active actin- and temperature-dependent contractile processes.

In the context of angiogenesis, a similar process also occurs and has been termed inverse blebbing \cite{Gebala2016}. Specifically, Ref.~\cite{Gebala2016} proposed a qualitative picture in which the hydrodynamic pressure gradient exerted by the blood flow on the apical aspect of an EC, promoted by local weakening of the cell actomyosin cortex, drives the nucleation of a spherical membrane invagination within the cell body. This protrusion is then mechanically inflated, while simultaneously F-actin polymerization is initiated on site and myosins-II motors are recruited from the cytoplasm to build a contractile actin cortex that fully envelopes and stabilizes the invaginated structure, thus promoting lumen expansion. This mechanism not only is reminiscent of cell blebbing with an inverted polarity (ordinary blebs are extracellular spherical protrusions of the cytoplasm \cite{Cunningham1995}), but also bears similarities with the formation of GVs in SC ECs (which are however typically ellipsoidal). 

Inspired by these analogies, here we formulate a quantitative model of GV formation that can reproduce all the qualitative features (i) – (iii). 
Our model shows that cell membrane is central to the mechanics of GV formation, in particular by enabling cells to limit the growth of the vacuoles, thus confirming previous similar conjectures \cite{Overby2011}. 
The model also predicts coarsening effects between multiple invaginating GVs, akin to Ostwald-ripening in phase separation \cite{lifshitz1961,voorhees1985,wagner1961,weber2019}, which are induced by their local competition for membrane stretching. 
The characteristic features (i) – (iii) and our own model predictions (see Discussion) can be relevant also in other experimental contexts where cells respond to large pressure gradients through inverse blebbing \cite{dumortier2019}.

\begin{figure*}[!tb]
\centering
\includegraphics[width=175mm,keepaspectratio]{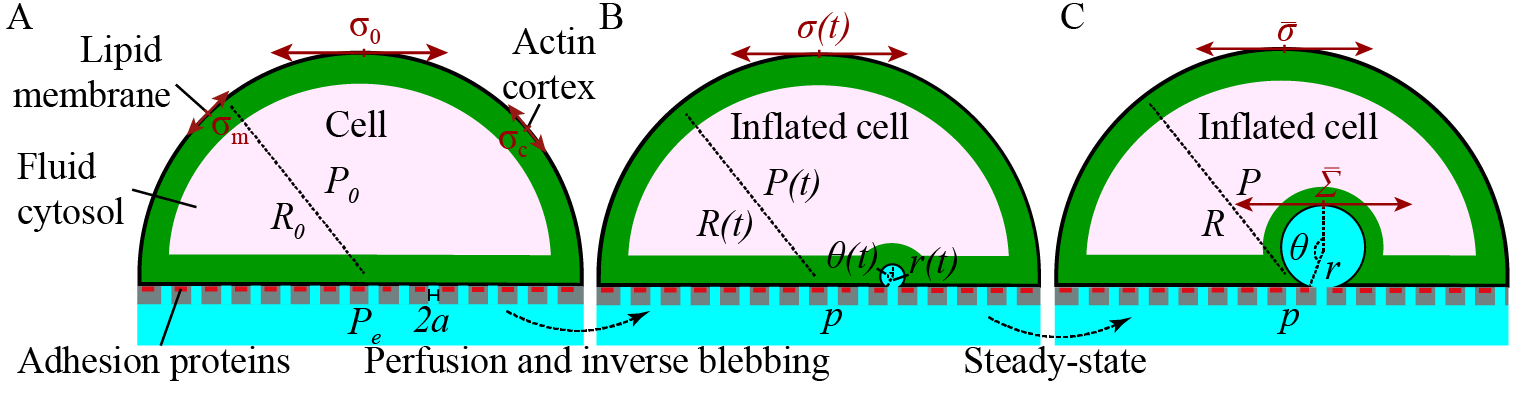}
\caption{
(A) Schematic of the cell mechanical model. A cell adheres to a filter that is immersed in a fluid at the pressure $P_e$. The cell is composed by the plasma membrane with tension $\sigma_{\rm m}$ , the actomyosin cortex with tension $\sigma_{\rm c}$  and the cytosolic fluid at the pressure $P_0\geq P_e$. The cell surface tension (membrane plus cortex) is $\sigma_0$. (B) Time dependent inverse bleb configuration. The inverse bleb is a spherical cap with radius r and opening angle $\theta$. Its overall surface tension is $\Sigma$. The cell maintains hemispherical shape with radius $R$, internal pressure $P$ and surface tension $\sigma$. (C) Stationary configuration after perfusion at constant pressure drop. For sufficiently long times (up to a few tens of minutes; comparable with the typical lifetime of giant vacuoles) the inverse bleb is enveloped by a contractile actomyosin cortex. The surface tensions of the cell and the inverse bleb, denoted as $\overline{\sigma}$ and $\overline{\Sigma}$ respectively, are equilibrated locally.  
}\label{figure1}
\end{figure*}

\section{Single-Cell Perfusion Model}

\subsection{Initial Setup and Cell Mechanical Model.} 

We consider an EC that adheres to a filter immersed in a fluid environment at the controlled pressure $P_e$ (Fig.~\ref{figure1} A). 
At first approximation, the cell at equilibrium assumes a hemispherical shape with radius $R_0$ due to its effective surface tension \cite{Boal2012}. 
The filter pores have width $2a$, which is consistent with the characteristic diameter of the meshwork pores of GVs \cite{Holmberg1959,Grierson1978}. 

To describe the cell, we consider the following mechanical model (Fig.~\ref{figure1} A): 
The cell is coated by a lipid membrane with in-plane surface tension $\sigma_{\rm m}$  and area expansion modulus $K_{\rm m}$ \cite{Boal2012}. 
While the former is due to purely entropic effects, the latter determines the elastic response of the membrane to mechanical stretch and is much larger in value (Table~\ref{table2}). 

Lying beneath the membrane and anchored to it, the cell possesses a contractile actin cortex, a dense cross-linked meshwork of actin filaments, myosin motors and actin-binding proteins \cite{Morone2006,Charras2006}. 
An in-plane tension $\sigma_{\rm c}$  is generated in the cortical network primarily by myosin-II motors \cite{Clark2011,Salbreux2012}. 
Prior to perfusion, the membrane is floppy and only the entropic surface tension contributes. 
We thus set the overall cell surface tension to be  $\sigma_0=\sigma_{\rm m} +\sigma_{\rm c}$ \cite{Clark2011}. 

The cell interior is filled with cytosolic fluid at the pressure $P_0\geq P_e$. Differently from previous models of ordinary cell blebbing \cite{Tinevez2009}, 
other internal elastic contributions, such as cytoskeletal components and cellular organelles, are neglected. 
In the present context, these only exert resistance to inverse bleb inflation, an effect that we will capture at a coarse-grained level (see below). 

The equilibrium configuration of the cell is determined by the Laplace law  $P_0-P_e=2 \sigma_0⁄R_0$.

\subsection{Inverse Blebbing and Perfused Configuration.} 

We can now perfuse the model cell by increasing the pressure of the ambient fluid inside the filter to $p\geq P_e$. 
We consider only constant pressure, but any time dependent protocol $p(t)$ can be studied using the same method. 

\begin{figure*}[!tb]
\centering
\includegraphics[width=175mm,keepaspectratio]{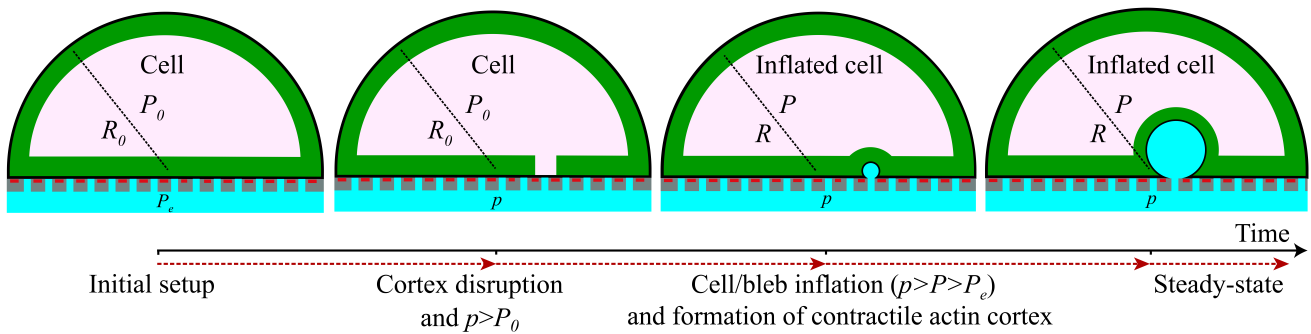}
\caption{
Schematic of inverse blebbing. See ref.~\cite{Gebala2016} for details.   
}\label{figure2}
\end{figure*}

This condition leads to the formation of an invagination of the cell basal aspect within the cellular body through inverse blebbing \cite{Gebala2016} (Fig.~\ref{figure2}): 
The hydrodynamic pressure drop $p-P_0$ across the cell basal surface together with a local disruption of the actomyosin cortex drives the nucleation of an intracellular membrane protrusion
\footnote{
In ref.~\cite{Gebala2016} it is shown that inflating inverse blebs do not express any F-actin or myosin II markers, 
thus demonstrating local cortex disruption at nucleation sites of inverse blebs.    
}. 
Both this inverse bleb and the cell then start inflating. Here we make the simplifying assumption that the cell inflates while keeping fixed its hemispherical shape. 
This implies that the cell adds new focal complexes instantaneously to cover its expanding basal surface (this is compatible with previous observations on spreading cells, see ref.~\cite{choi2008}). 

Simultaneously, actin polymerization and myosin-II motors recruitment are initiated on the inverse bleb surface, such that a contractile actin shell, akin to the cell cortex, is formed around the inverse bleb. 
With time the shell grows to the same thickness of the cell cortex, such that its surface tension becomes equal to the cortical tension. 
For sufficiently long times the system can reach mechanical equilibrium, where the surface tensions of both the contractile shell and the plasma membrane balance the pressure force exerted at the cell-bleb interface. 

This mechanism is reminiscent of ordinary cellular blebbing \cite{Cunningham1995}. 
The crucial difference is that inverse blebs have opposite polarity, in that they inflate inside the cellular body. In addition, while blebs are self-sustained through the cytoplasmic pressure, inverse blebs require an external pressure drop.

We assume that the inverse bleb has spherical cap shape with radius $r$ and opening angle $\theta$ $(\pi/2\geq \theta \geq \pi)$ 
\footnote{
For inverse blebs with $0<\theta<\pi⁄2$, the total volume is $\lessapprox (2\pi a^3)/3\approx 0.03$ $\mu\text{m}^3$. 
This estimate is obtained by using the value of $a$ reported in Table~\ref{table2}. 
This was measured in ref.~\cite{Grierson1978} at the physiological pressure drop $\Delta p=7$ mmHg. 
The argument holds also at larger pressures; e.g., at $\Delta p=14$ mmHg  Grierson and Lee measured $2a\approx 0.9$ $\mu\text{m}$ \cite{Grierson1978}, 
such that $(2\pi a^3)/3\approx 0.19$ $\mu\text{m}^3$. At these pressures, they measured GVs with volume about $36.45$ $\mu \text{m}^3$. 
Given their much smaller size compared to true GVs (e.g. GV volume=15.67 $\mu m^3$ at $\Delta p=7$ mmHg  according to the data from ref.~\cite{Grierson1977}), 
we neglect these solutions. 
}
, which are related by the geometrical condition $r\sin{(\theta)}=a$. 
The instantaneous surface tension of the inverse bleb is denoted by $\Sigma$. 
Likewise, the cell maintains its initial hemispherical shape but with radius $R$, internal pressure $P$ and surface tension $\sigma$ (Fig.~\ref{figure1} B). 
We remark that at nucleation $\Sigma=\sigma_{\rm m}$, because cortical disruption (see above) causes a local reduction of the surface tension.

\subsection{Modeling Inverse Bleb Growth.} 

To enable the inflation of the inverse bleb within the cellular body, the pressure inside the inverse bleb $p$ must be larger than the intracellular pressure $P$ \cite{Safran2018}. 
The membrane mechanical behavior is typically described in terms of the fluid-mosaic model \cite{Singer1972}. 
This model prescribes that the membrane can be regarded to all effects and purposes as a viscous fluid. This implies that the membrane tensions of the cell and the inverse bleb should be able to equilibrate globally even at short times (e.g., a few seconds). Considering that actin cortical dynamic is also fast (see Table~\ref{table2}) for both the cell and the inverse bleb, their surface tensions all together (membrane plus cortex contribution) should also be able to equilibrate globally even at short times. 

This expectation, however, has been refuted by recent experiments of membrane tethering. 
These experiments have demonstrated that the surface tension of the inverse bleb and the cell can only equilibrate locally for short to intermediate timescales (up to about ten minutes) \cite{shi2018}. 
This counterintuitive effect is caused by the impediment to tensional flow exerted by cytoskeleton-bound transmembrane proteins 
\footnote{
In ref.~\cite{shi2018} the relaxation dynamics of membrane tension was studied by pulling tethers at fixed distance between one another and by monitoring their fluorescence, as a measure of surface tension, in response to a perturbation of one of the two tethers. These experiments demonstrate that in intact cells, for tethers as close as $5$ $\mu \text{m}$, no correlation between the fluorescence signals are observed over up to $500$ s. The two tethers are thus effectively independent at this timescale. Conversely, when measurements are repeated in cellular blebs, the signals are strongly correlated (lag time $< 1 s$). In this case, the tension almost instantaneously equilibrates in the tethers, in agreement with the fluid-mosaic model  \cite{Singer1972}.
}. 
As a result, for timescales up to a few tens of minutes, the surface tensions of the cell and of the inverse bleb can be different. 
Over these timescales, the inverse bleb tension equilibrates within a membrane patch of radius $d\approx10 \, \mu m$  (see Table 2). 

For longer timescales, the inverse bleb and the cell interact through the slow lipid flow that is driven by the tensional gradient established at their boundary \cite{shi2018}. 
This enables equilibration of surface tension throughout the entire system at long times. 
However, because the characteristic lifetime of GVs is of the order of a few tens of minutes, we ignore the eventual equilibration and consider their tension separately (Fig.~\ref{figure1} C).

\section{Inverse bleb radius dynamic: The Rayleigh-Plesset equation}

Given the above discussion, we now formulate a dynamical model for GVs. 
The key ingredients are: 
\begin{enumerate}
\item the inverse bleb growth dynamic is overdamped \footnote{Assuming cytoplasmic density $\rho\approx 1 \, \text{g/cm}^3$ \cite{moran2010} and viscosity $\mu\approx 0.2-0.3 \, \text{Pa}\cdot\text{s}$ \cite{Chugh2017}, characteristic GV length $L\approx 5-100 \, \mu\text{m}$ \cite{Pedrigi2011} and bleb growth rate $U=L⁄(60\,\text{s})\approx (0.08-2) \, \mu\text{m/s}$, we obtain the estimate for the Reynolds number $\text{Re}=\rho U L⁄\mu \lessapprox 10^{-6}$. Characteristic length and velocity of GVs are estimated according to the data in vitro from ref.~\cite{Pedrigi2011}. As these vacuoles are typically much larger than those observed in vivo \cite{Grierson1977}, this calculation overestimates the Reynolds number. Therefore, we expect this assumption to hold also in physiological conditions.} and is driven by the mechanical pressure, viscous resistance and surface tension forces that are exerted uniformly at the cell-bleb interface. 
\item The intracellular pressure $P$ is fast equilibrating, i.e., its rate of change is negligible compared to the corresponding rates of the tensions and the radii. 
\item Finally, there is no leakage of intracellular material to the extracellular environment during inverse blebbing. This condition is experimentally supported \cite{Pedrigi2011} and prescribes volume conservation at all times. 
Consequently, the cell radius can be specified by the geometrical relation $R^3=R_0^3+2r^3 (2+\cos{\theta})\left[\sin{(\frac{\theta}{2})}\right]^4$. The volumetric contribution of the basal pore can be neglected (SI Appendix Sec. I).
\end{enumerate} 

Taken all altogether, the above assumptions led us to model the dynamic of the radius of the inverse bleb in terms of the Rayleigh-Plesset equation without the inertial terms \cite{rayleigh1917,plesset1949,plesset1977}:
\begin{equation}
\frac{\mathrm{d} r(t)}{\mathrm{d} t}=\frac{r(t)}{4\nu} \left[p-P(t)-\frac{2\Sigma(t)}{r(t)} \right] , 
\label{eq:RL}
\end{equation}
where the effective viscosity $\nu$ describes the resistance to inverse bleb growth exerted by the inner cell body. This equation expresses balance of the mechanical force due to the bleb surface tension, and of the mechanical pressure and viscous forces exerted by the inner cell body at the cell-bleb interface. The intracellular pressure is set by the mechanical equilibrium prescribed by the Laplace law for the cell $P(t)-P_e=2\sigma(t)⁄R(t)$ (Fig.~\ref{figure1} B). The surface tensions $\Sigma$ and $\sigma$ are functions of the relative surface area strains of the inverse bleb and cell respectively (see below). As such, they are functions of the respective radii,  $r$ and $R$. In turn, these specify their time dependence. Complemented by volume conservation, which relates the variables $R$ and $r$, and by the geometrical relation between $\theta$ and $r$, these equations only depend on the independent variables $(r,\Sigma,\sigma)$ and can be solved numerically for fixed pressure drop $\Delta p= p-P_e$. 

In particular, the steady-state solutions of these equations (Fig.~\ref{figure1} C) are determined by
\begin{equation}
\Delta p=\frac{2\overline{\Sigma}(r)}{r}+\frac{2\overline{\sigma}(r)}{R}
\label{eq:steady-state}
\end{equation}
We stress that this limit describes the system configurations for timescales close to the characteristic lifetime of GVs, 
but small enough that the lipid flow established at the inverse bleb-cell interface can be neglected. 
The equation is closed and can be solved for $r$ in terms of $\Delta p$. 
We note also that correction terms due to membrane bending rigidity are expected to be negligible (SI Appendix Sec. II).

\section{Surface tension dynamic: Dependence on the Relative Area Strain} 

The surface tensions of both the inverse bleb and the cell (considered separately; see above) follow a relaxation dynamic toward their value at steady-state, $\overline{\Sigma}$ and $\overline{\sigma}$ respectively. 
In our context, both this target tension and the characteristic timescale of relaxation are determined dynamically during the growth process through a dependence on the relative area strains, 
$\epsilon_{\rm B}$  for the bleb and $\epsilon_{\rm C}$  for the cell. 
We define $\epsilon_{\rm C}$ as $\epsilon_{\rm C}(t)=[A(t)-A_0]⁄A_0$ 
with $A_0=3\pi R_0^2-\pi d^2$ 
and $A=3\pi R^2-\pi d^2$; 
$\epsilon_{\rm B}$ as 
$\epsilon_{\rm B}(t)=[S(t)-S_0]⁄S_0$ 
with $S_0=\pi d^2$ and 
$S=2\pi r^2 (1-\cos{\theta})+\pi(d^2-a^2 )$. 
We assume two mechanical regimes as a function of these relative area strains.

\subsection{Cortex-dominated regime} 

The plasma membrane can adapt to possibly large relative area strains at constant surface tension (up to $\epsilon^*=100$ at cellular level
\footnote{Larger cellular strains (up to $1000\%$) have also been observed in the experiments of doming endothelial monolayers of ref.~\cite{Latorre2018}. 
However, cellular strains larger than about $100\%$ are a manifestation of a super-elastic response \cite{wayman1998}, which we neglect in the present context. }; see \cite{Latorre2018}). 
Different pathways have been identified that enable cells to buffer the surface area required \cite{gauthier2012}. 
The membrane possesses different types of lipid reservoirs, e.g., caveolae \cite{Sinha2011,Figard2016}, 
membrane wrinkles and invaginations \cite{Hallett2007,Goudarzi2017}, 
microvilli and other membrane protrusions \cite{Figard2013,Figard2014}, 
that can be flattened out. 
Additional membrane can also be supplied by active exocytic processes within timescales of seconds to minutes \cite{kosmalska2015}, 
thus comparable with the typical lifetime of GVs (see above). 
The dominant contribution to the steady-state tension in this regime is provided by the actomyosin cortex for the cell and by the actin contractile shell for the inverse bleb. 
As such, $\overline{\Sigma}=\sigma_0$ and $\overline{\sigma}=\sigma_0$. 
The cell cortex is already initialized at its steady-state value. 
Therefore, as long as the cell remains in this regime, its surface tension does not change in time, i.e., $\sigma(t)=\overline{\sigma}=\sigma_0$. 
The contractile shell enveloping the inverse bleb must be rebuilt instead following the local disruption of the cortex occurring at nucleation. 
Correspondingly, tensional dynamic is governed by the characteristic timescale $\tau_{\rm c}$  of actin turnover and myosin recruitment \cite{Charras2006}. 
Assuming exponential relaxation, we can write 
\begin{equation}
\frac{\mathrm{d} \Sigma(t))}{\mathrm{d} t}=\frac{1}{\tau_c}[\sigma_0-\Sigma(t)].
\label{eq:Sigma}
\end{equation}

\subsection{Membrane-dominated regime} 

Once the membrane reservoirs are depleted, which corresponds to the threshold relative area strain $\epsilon^*$, 
no additional lipids can be added to the plasma membrane, 
such that further increase in the surface area must be obtained by mechanically stretching the membrane. 
This mechanical process induces abrupt increase of the surface tension due to the large area expansion modulus of the membrane \cite{Boal2012}. 
Therefore, the dominant contribution to the surface tension in this regime is given by the taut membrane. 
Furthermore, in these conditions the membrane is in the two-dimensional liquid-like state \cite{Singer1972}, 
such that its equilibration can be regarded as an instantaneous process. 
In our context, this regime is only relevant for the inverse bleb. 
In fact, when only one GV is formed, the area strain of the cell never reaches the threshold $\epsilon^*$ (up to at least 30 mmHg). 
Therefore, we set $\Sigma(t)=\overline{\Sigma}(t)=\sigma_0+K_{\rm m} e^{[(S(t)-S^*)⁄S^*)]}$ with $S^*=S_0 (1+\epsilon^*)$.

\begin{table}[!tb]
\centering
\begin{tabular}{l@{\hskip 0.25in}l@{\hskip 0.25in}l@{\hskip 0.25in}}
Parameter & Description & Units \\
\toprule
$P_e$ & Ambient pressure & mmHg \\ 
$R$ & Cell radius & $\mu$m \\  
$2a$ & Pore diameter & $\mu$m \\ 
$\sigma$ & Cell surface tension & pN\/$\mu$m \\  
$K_m$ & Area expansion modulus & N\/m \\  
$p$ & Pressure inside the filter & mmHg \\  
$r$ & Inverse bleb radius & $\mu$m \\  
$\theta$ & Inverse bleb opening angle & adim \\  
$\Sigma$ & Inverse bleb surface tension & pN\/$\mu$m \\  
$P$ & Intracellular pressure & mmHg \\
$\Delta p$ & Perfusion pressure drop & mmHg \\  
$\epsilon$ & Relative area strain & adim \\  
$A$ & Cell surface area & $\mu\text{m}^2$ \\
$S$ & Inverse bleb surface area & $\mu\text{m}^2$ \\ 
$\tau$ & Characteristic timescale & s \\  
$\nu$ & Effective cellular viscosity & $\text{pN} \cdot \text{s}/\mu\text{m}^2$ \\
\botrule
\end{tabular}
\caption{Notation used in the model. We further use subscripts “0” to denote quantities corresponding to the initial cellular configuration, “m” and “c” to denote quantities corresponding to the membrane or the actomyosin cortex respectively, “B” and “C” to denote quantities corresponding to the inverse bleb or the cell respectively. The superscript “*” denotes thresholds.  }\label{tab:parameters}
\end{table}

\section{Results}
  
\subsection{Inverse blebs dynamics} 

\begin{figure}[!tb]
\centering
\includegraphics[width=75mm,keepaspectratio]{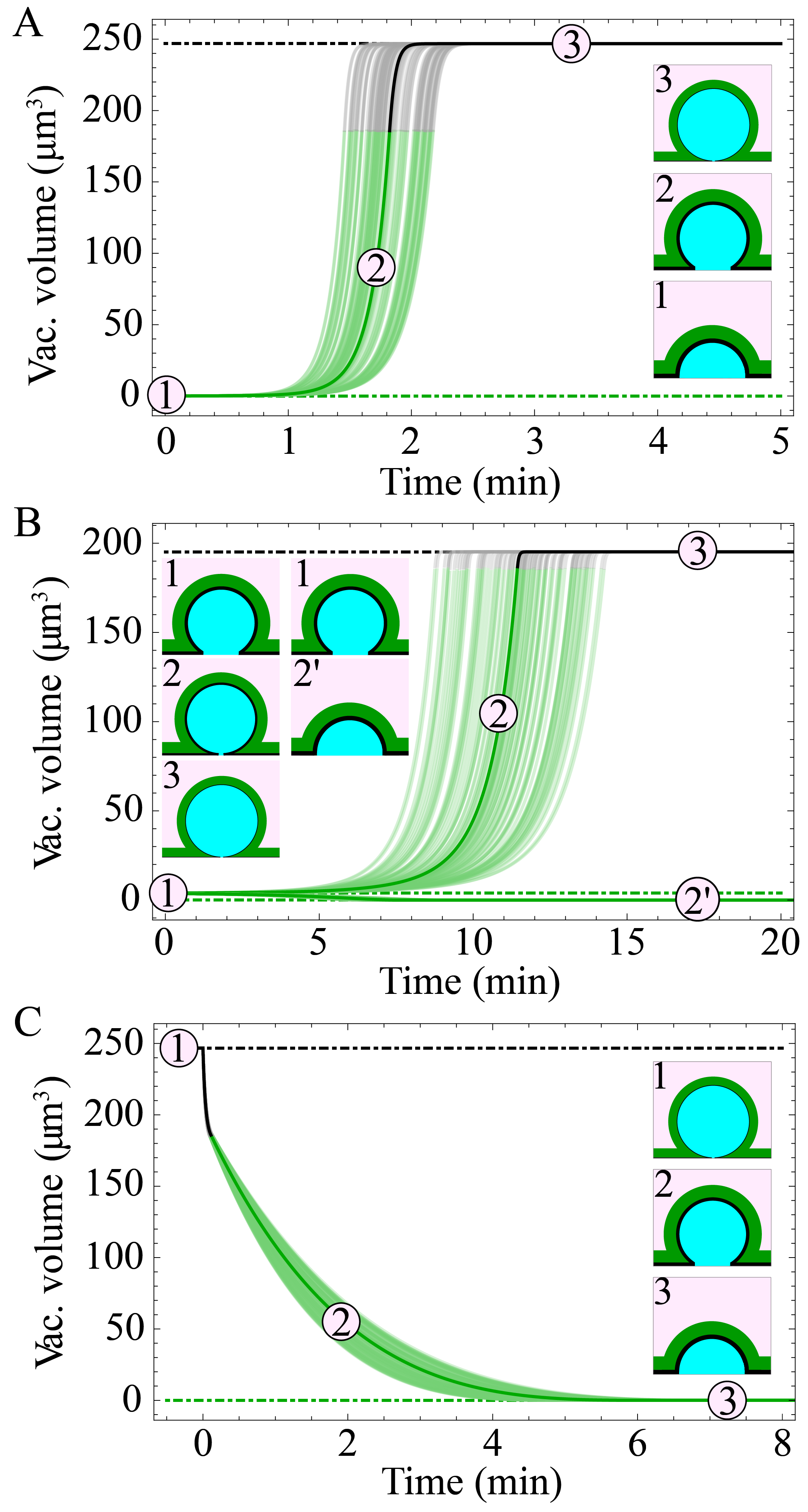}
\caption{Dynamics of inverse blebs. 
Model parameters are chosen according to Table 2, in particular $R_0=10$ $\mu$m, $d=10$ $\mu$m and $\epsilon^*=0.5$. 
The pressure drop is kept constant for all times.  
Blurred regions correspond to trajectories obtained by allowing for up to $20\%$ variability in the dynamical parameters $\tau_c$ and $\nu$. 
(A) Large pressure drops. We set here $\Delta p=30$ mmHg. 
The giant vacuole is initially nucleated with $r=a$ and $\Sigma=\sigma_{\rm m}$. 
At this pressure, the inverse bleb inflates until it reaches the target giant vacuole configuration in the membrane-dominated regime (within about 2 min). Then growth stops. 
(B) Physiological pressure drops. We set here $\Delta p=7$ mmHg. 
The dynamic become sensitive on the nucleation conditions. Different nucleation radii can lead to either inverse bleb shrinkage or growth. 
(C) Giant vacuole collapse. The giant vacuole is initially set in its steady-state configuration (at $\Delta p=30$ mmHg). 
At time $t=0^+$ the pressure drop is removed. The giant vacuole slowly deflates (within about 5 min).   
}\label{figure3}
\end{figure} 

Numerical analysis of the model equations enables us to investigate the dynamical features of inverse blebbing. 
We adopt the following protocol: 
The pressure drop $\Delta p$ is established at the initial time $t=0$ and kept constant throughout the growth process. 
The inverse bleb is initially invaginated with $\Sigma=\sigma_{\rm m}$.
This condition replicates the local disruption of the actomyosin cortex observed during inverse bleb nucleation \cite{Gebala2016}. 

We observe three distinct qualitative scenarios: 
For sufficiently small $\Delta p$, the inverse bleb immediately deflates independently on its radius at nucleation (not shown). 
By construction, our model can only predict the deflation dynamic of the inverse bleb until $r=a$. 
This result suggests the existence of a threshold pressure for GV formation, $\Delta p^*$. 
For large (and rather non physiological) pressure drops the inverse bleb grows toward a target configuration within a characteristic timescale of the order of minutes. 
Then, growth stops (Fig.~\ref{figure3} A). 
This target configuration corresponds to the steady-state GV described by Eq.~\eqref{eq:steady-state}. 
The growth process in this case is again independent on the inverse bleb radius at nucleation. 
For intermediate pressure drops, the dynamic of the inverse bleb becomes sensitive on the nucleation radius. 
For sufficiently small radii, inverse blebs shrink; 
conversely, for large enough radii, inverse blebs grow (Fig.~\ref{figure3} B). 
This result suggests the existence of a threshold nucleation radius for GV growth at fixed physiological pressure drops.  

We can also study GV collapse (Fig.~\ref{figure3} C). The GV is initialized at its steady-state configuration for pressures $\Delta p>\Delta p^{\dagger}$ (here $\Delta p=30$ mmHg). 
At time $t=0^+$ the pressure drop is removed and the GV let evolve according to Eqs.~\eqref{eq:RL} and~\eqref{eq:Sigma}. 
The model predicts shrinkage of the GV within a larger timescale than the corresponding timescale for growth at the same pressure (here about 4-6 min). 
Moreover, we observe an even longer timescales for GV collapse, if the cortical tension is reduced (SI Appendix Fig. S3). 

\subsection{Inverse blebs steady-state configurations} 

\begin{figure}[!tb]
\centering
\includegraphics[width=75mm,keepaspectratio]{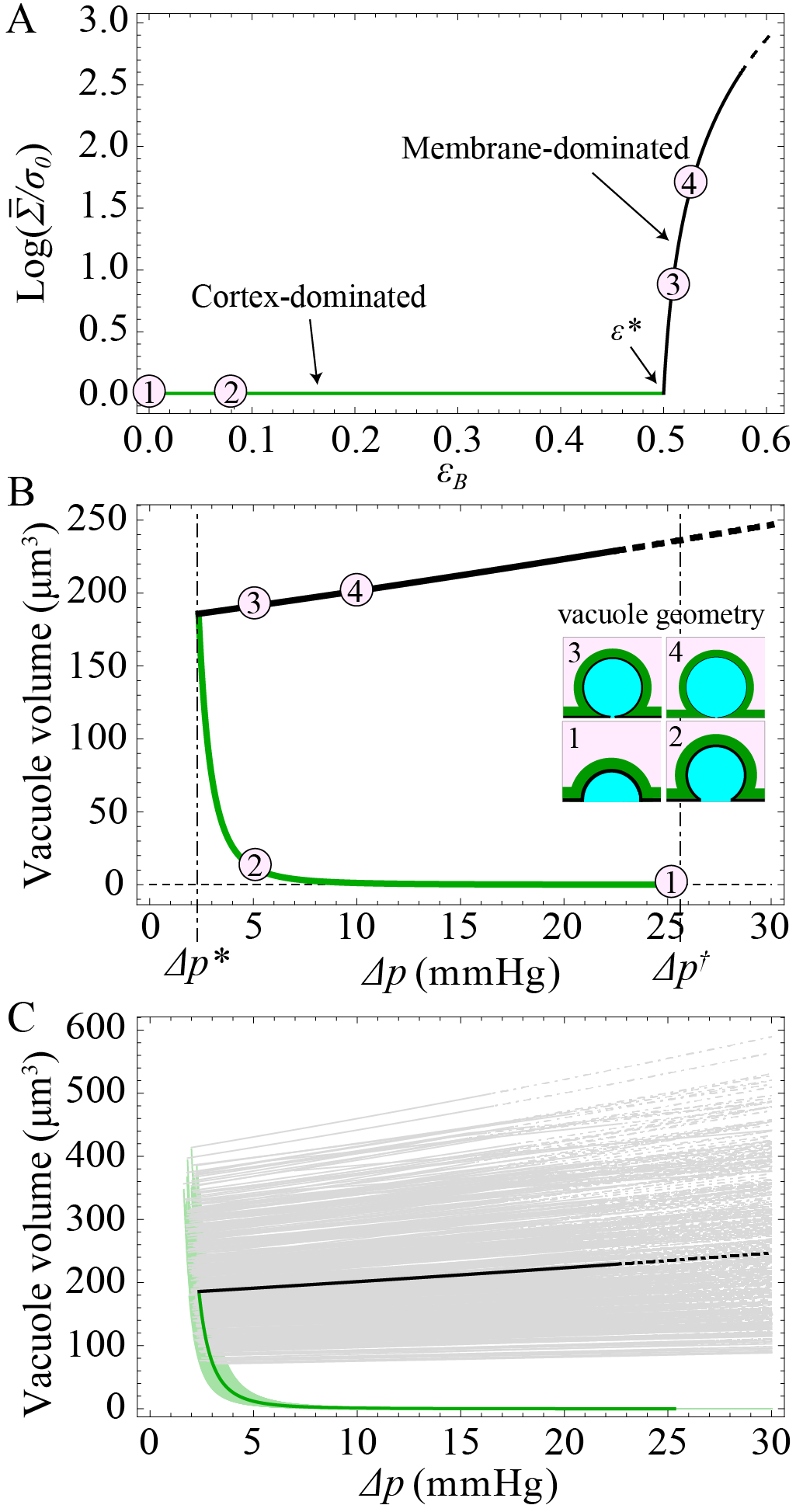}
\caption{Steady-state configurations of giant vacuoles. 
(A) Dependence of the surface tension of the vacuole at steady-state $\overline{\Sigma}$ on its area strain $\epsilon_{\rm B}$. 
The two mechanical regimes are highlighted with different colors (green/black respectively). 
We assume the same functional form for the cell. 
(B) Vacuole volume vs. perfusion pressure drop $\Delta p$. 
Examples of vacuole configurations are plotted to illustrate the dependence of the geometrical shape on the pressure drop. 
Characteristic parameters in panels A and B are chosen according to Table 2. 
For $\Delta p$ large, the surface tension can reach lytic values (here corresponding to area strains $5\%$ $\epsilon^*$), 
when rupture of the plasma membrane occurs (dot-dashed line). 
(C) Stationary configurations obtained by allowing for up to $20\%$ variability in the characteristic parameters of the model. 
We plot simulation results for 500 different sets of parameters chosen randomly. 
These results indicate that single-cell predictions are qualitatively robust against physiological tissue variability.  
}\label{figure4}
\end{figure} 

Numerical analysis of the steady-state equation (2) confirms the previous results (Fig.~\ref{figure4} B). 
For large enough pressures $(\Delta p \geq \Delta p^{\dagger})$ only one steady-state configuration exists. 
This explains why inverse blebs grow to this GV configuration independently on the nucleation radii at these pressures. 
For intermediate pressures $(\Delta p^* \leq \Delta p \leq \Delta p^{\dagger})$ two steady-state configurations exist with $\overline{\Sigma}$ either in the cortex- or in the membrane-dominated regime. 

Using linear stability analysis, we can study how steady-state inverse blebs respond to small perturbations of their shape. 
We find that inverse blebs in the membrane-dominated regime are stable, while those in the cortex-dominated regime are unstable (SI Appendix Sec. III). 
The dynamical picture becomes clear: 
If the inverse bleb is nucleated with a radius smaller than the radius of the steady-state configuration in the cortex-dominated regime, it will shrink. 
Conversely, if it is nucleated with a larger radius, it will grow toward the steady-state GV configuration in the membrane-dominated regime. 
The threshold nucleation radius for GV growth at fixed intermediate pressures is therefore of the order of the radius of the steady-state configuration in the cortex-dominated regime. 
We remark that cells can ensure stability of the GVs in the cortex-dominated regime by enveloping them in additional contractile actomyosin structures (like actin rings) 
that respond elastically (at short times) to the perturbation \cite{Tinevez2009} (SI Appendix Sec. III).   

These considerations also yield an estimate of the GV nucleation pressure, $\Delta p^*\approx 2\sigma_0 \frac{(R+r)}{Rr}$, 
where $R$, $r$ are both $\epsilon^*$ -dependent. 
This function decreases monotonically as $r$ increases (due to volume conservation) independently of the specific geometry of the perfused system.  
As such, $\Delta p^*$ only depends on the membrane reservoir size through the parameter $\epsilon^*$  and on the initial cell surface tension $\sigma_0$ (linearly). 

All these results are qualitatively robust against physiological tissue variability. 
This is confirmed by numerically solving equation~\eqref{eq:steady-state} while allowing for up to $20\%$ change in the model parameters (chosen randomly). 
All stationary configurations obtained do not differ qualitatively between themselves and from the reference case (Fig.~\ref{figure4} C). 
We remark that the robustness of our model results against variability of the parameter $2a$ (the pore diameter) is especially relevant, 
because in physiological tissues SC ECs adhere to a discontinuous basement membrane possessing a broad distribution of pore sizes.

\subsection{Coarsening of Multiple Giant Vacuoles through Ostwald Ripening} 

In physiological conditions multiple GVs typically invaginate simultaneously. 
As a proxy of this general situation, we study the case of two growing GVs. 
Neglecting steric interactions between the blebs and contact-driven shape deformations, 
Eqs.~\eqref{eq:RL} and ~\eqref{eq:Sigma} are formally valid for both the cell and each inverse bleb. 
However, the volume conservation must now account for the contribution of all blebs; 
therefore, the relation determining the cell radius becomes 
$R^3=R_0^3+2\sum_{i} r_i^3 (2+\cos{\theta_i})[\sin{(\theta_i⁄2)}]^4$ $(i=1,2).$      

\begin{figure*}[!tb]
\centering
\includegraphics[width=175mm,keepaspectratio]{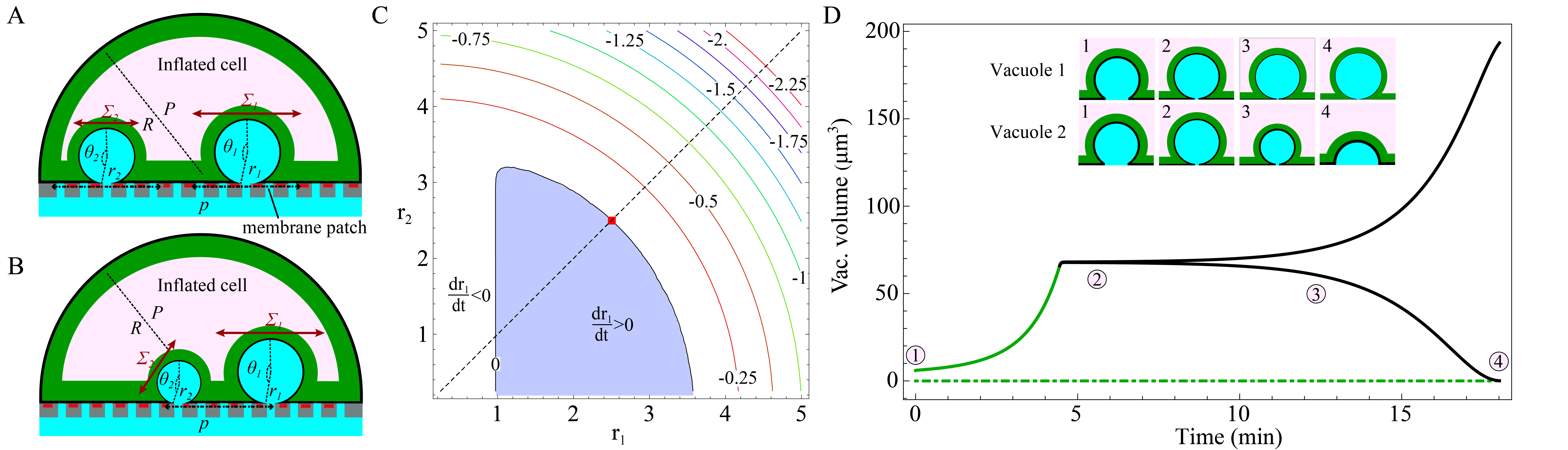}
\caption{Coarsening process between multiple invaginating giant vacuoles. 
(A) Two vacuoles invaginate simultaneously at a distance larger than $d$. 
Because they do not share the same equilibrating membrane patch, they grow independently. 
(B) Two vacuoles invaginate simultaneously at a distance smaller than $d$. 
The vacuoles share the same equilibrating membrane patch. 
Their growth processes are no longer independent. 
(C) Contour plot of $\mathrm{d} r_1⁄\mathrm{d}t$ vs. $(r_1,r_2)$ at steady-state. 
The pressure drop is set at physiological values, here $\Delta p=7 \text{mmHg}$. 
Other model parameters are set according to Table 2, in particular $R_0=10$ $\mu$m , $d=10$ $\mu$m and $\epsilon^*=0.5$. 
The structure of contours is similar to the standard drop coarsening picture as in the Ostwald ripening scheme. 
(D) Time series of the volume of two giant vacuoles illustrating the coarsening dynamic. 
Model parameters are chosen as in panel C. 
The two giant vacuoles are initialized simultaneously at time $t=0$ with different, randomly chosen, initial sizes close to their steady-state configuration in the cortex-dominated regime, 
here $r_1\approx 1.1238$ and $r_2\approx 1.1241$, and $\Sigma_i=\sigma_{\rm m}$.   
}\label{figure5}
\end{figure*}

We can now distinguish two different qualitative scenarios: 
First, the GVs can be nucleated at a distance larger than the characteristic length-scale $d$ (see above), 
such that they grow by increasing the surface area of two different membrane patches (Fig.~\ref{figure5} A). 
These vacuoles are therefore fully independent, 
and their steady-state and dynamical properties are the same as those of a single GV. 

Second, the GVs can be nucleated at a distance smaller than $d$, 
meaning that they grow by sharing the same membrane patch (Fig.~\ref{figure5} B). 
The two vacuoles are therefore no longer independent: 
Their surface tensions converge to the same target tension $\overline{\Sigma}$ dependent on the relative area strain $\epsilon_B$ with 
$S=\sum_i r_i^2  (2+\cos{\theta_i}) [\sin{(\theta_i⁄2)}]^4+\pi (d^2-2a^2)$. 
The stationary configurations predicted by the model do not change qualitatively from the single-inverse bleb picture: 
The second bleb only reduces the amount of stored membrane locally available for each inverse bleb. 
Consequently, the threshold pressure for GV nucleation $\Delta p^*$ increases and the maximal size of each vacuole decreases.             

When we consider the dynamics of the vacuoles, we observe a coarsening mechanism akin to Ostwald-ripening. 
This is demonstrated by the contour plot of $\mathrm{d} r_1⁄ \mathrm{d}t$ as a function of $r_1$ and $r_2$ (Fig.~\ref{figure5} C) at steady-state (i.e., $\Sigma_1=\Sigma_2=\overline{\Sigma}$ and $\sigma=\overline{\sigma}$). 
The structure of the contours is similar to the standard drop coarsening picture as in the Ostwald ripening scheme. 
In particular, focusing on the equal size configuration $(r_1=r_2)$ at the zero-contour line (indicated by a red square), 
fluctuations (e.g., increasing $r_1$ while decreasing $r_2$) around that point can generate an instability 
that leads to the growth of the bigger GV and the shrinking of the smaller one. 

The dynamic of the coarsening process can also be characterized (Fig.~\ref{figure5} D): 
The two GVs, nucleated with different sizes, grow until their surface tensions reach the membrane-dominated regime. 
In this condition, two scenarios can occur: 
If the two vacuoles are small, the cell can fully buffer the area increase required through its membrane reservoirs. 
The two vacuoles therefore grow independently on one another. 
Conversely, if the two vacuoles are large enough to exhaust the cellular membrane reservoirs, 
they start competing between one another in order to stretch locally the lipid membrane 
and thus increase their own surface area. 
For long times symmetry breaking occurs: 
Only one GV succeeds in expanding its surface area and reaching the stationary configuration, 
whereas the other vacuole shrinks. 
In physiological conditions such symmetry breaking can be triggered by either different nucleation conditions among the vacuoles 
or by shape fluctuations that can be due, among others, to natural inhomogeneities of the inverse bleb growth dynamics 
(these effects are however neglected in our model).

\section{Discussion}

In summary, by identifying GV formation as inverse blebbing, 
we formulated a biophysical model of this process that can recapitulate all the characteristic morphological and dynamical features of GVs (i) – (iii). 
The model thus elucidates the mechanisms underlying the dynamics of GVs at coarse-grained level: 
Upon nucleation, both the GV and the cell adjust their configurations to reach mechanical equilibrium locally. 
Vacuole and cell surface tensions are modulated on their relative areal strains according to the following qualitative picture (Fig.~\ref{figure4} A): 
For relative strains smaller than the threshold $\epsilon^*$ (corresponding to the size of the local membrane reservoir), 
the surface tension is in the cortex-dominated regime where all area increase is buffered by membrane reservoirs at constant tension. 
As shape changes are accommodated by active remodeling of the cell and bleb actomyosin cortices, 
tensional dynamic in this regime is governed by the characteristic timescale of actin turnover and myosin-II recruitment \cite{Charras2006}. 
Conversely, when these reservoirs are depleted, the surface tension enters the membrane-dominated regime 
where the area increase is buffered by the local mechanical stretch of the membrane. 
This mechanism induces an abrupt increase of the surface tension. 
Tension dynamic in this regime is governed by the characteristic timescale of membrane stretching, which is assumed to be a fast process. 
To all effects and purposes, we can consider it as instantaneous. 
For the case of a single GV, the picture simplifies. 
The cell in fact never enters the membrane-dominated regime (we verified this numerically up to pressures about 30 mmHg). 
Its surface tension is thus equal to $\sigma_0$ at all times during the lifetime of the GV.      

Our results elucidate the multiscale complexity underlying the GV growth process: 
In response to externally applied macroscopic stresses (in the context of Schlemm’s canal, the hydrodynamic pressure), 
several microscopic, active and passive, cellular processes are activated, including actin turnover, myosin-mediated contractile force generation and unfolding of membrane reservoirs, 
which drive the mechanical shape change that the cell undergoes. 
In turn, these processes are self-regulated at the macroscale by the surface tension. 
In this respect, our results particularly highlight the role played by the membrane in determining the existence and stability of GVs at physiological pressures. 
Our analysis also shows that the characteristic timescale induced by the viscous resistance exerted by the cell body on the inverse bleb is much larger than the characteristic timescale of actin turnover. 
While the latter is important to determine the growth characteristics at short times after nucleation, 
the former specifies the long time behavior. 
In the context of glaucoma, our results challenge the established idea that cellular stiffness is the main determinant of GV formation, 
thus confirming previous phenomenological suggestions \cite{Overby2011}.

Our model further reveals that GVs are stabilized by the elastic response of the actomyosin cortex enveloping them upon reaching the steady state. 
This result indirectly specifies the possible pathways for GV collapse: 
GVs can shrink either if an apical pore is formed thus constituting, together with the meshwork pore, a transcellular channel through which aqueous humor can flow; 
or if an active contractile force is exerted on the vacuole by actin contractile structures additional to the cortical shell; 
or through coarsening (see below).     

We remark that in our simulations we accounted for membrane reservoir sizes up to $100\%$ 
as recently observed in experiments on doming three-dimensional epithelia \cite{Latorre2018}. 
However, these same experiments have discovered examples of ECs with values of relative areal strain up to $300\%$ 
(thus much larger than those currently accounted for in our work). 
These extreme deformations have been interpreted as manifestations of super-stretched cellular states. 
Our coarse-grained dynamical model holds formally also in this case, 
but the target tensions $\overline{\Sigma}$ and $\overline{\sigma}$ must account for the strain softening induced by cortical dilution at high-strains underlying active super-elasticity \cite{Latorre2018}. 
We neglected this mechanism here as it is irrelevant for GV formation.

\section{Model predictions} 

The model provides the following novel predictions on GV dynamics:

\begin{enumerate}[label=\Roman*.]

\item{There exists a threshold pressure $\Delta p^*$ for GV formation. 
This pressure depends only on $\epsilon^*$ and on the cortical tension $\sigma_c$ (through the cell initial surface tension $\sigma_0$) 
and is independent of GV geometry. 
In details, $\Delta p^*$ decreases either if $\sigma_c$ is reduced or if $\epsilon^*$ is increased. 
Conversely, if $\sigma_c$ is increased or $\epsilon^*$ decreased, $\Delta p^*$ increases. 
Moreover, an inverse relation between $\Delta p^*$ and the vacuole sizes is shown: 
Decreasing $\Delta p^*$ induces larger GVs in the membrane-dominated regime. 
These predictions suggest a number of possible validating experiments: 
ECs could be treated with myosin-impairing drugs (such as Blebbistatin) to reduce the contractility of the cortex and thus $\sigma_c$; 
alternatively, lipids could be added ad hoc to increase the membrane reservoir pool. 
In both cases, our model predicts increased GV sizes, a result that could be assessed via experimental measurements.}

\item{At physiological pressure drops, there exists a threshold nucleation radius for GV growth of the order a few microns (about 1 $\mu$m at 7 mmHg).}   

\item{The timescale required for GV collapse, upon removal of the pressure drop, is larger than the corresponding timescale for GV growth at the same pressure conditions. 
Moreover, the model predicts that the characteristic timescale for GV collapse becomes longer if the cortical tension $\sigma_c$ is reduced. 
Experimental evidence validating this prediction has been recently found (cit. and/or private communication). }

\item{There exists a coarsening mechanism between multiple GVs inflating within the same cell, 
which is driven by the competition between these vacuoles for locally stretching the membrane. 
This generates Ostwald ripening: 
For long times symmetry breaking occurs and only one vacuole can grow towards the stationary configuration at the expense of all the other ones that instead shrink. 
As this effect is strictly related to the inter-bleb competition for membrane, 
its experimental verification could be used as a proxy to test the role of the cell membrane during GV formation.  }

\end{enumerate}

\section{Model limitations} 

We formulated here a coarse-grained biophysical model of GV formation as inverse blebbing. By its own nature, our model can only capture the macroscopic effects induced by the microscopic cellular processes involved in GV formation. This approximation captures the long-time morphological and dynamical features of GVs, but it fails to capture the characteristic processes occurring on short timescales. In this respect, our model does not contain any explicit elastic response contribution of the cell body on the inverse bleb (in fact, this term can also be shown to be negligible with respect to the effective viscous response). These short-time processes are relevant, particularly, to analyze the initial nucleation of GVs, which, at the level of the current discussion, only enters the theoretical description as a boundary condition. However, we believe the current regime of focus is most relevant experimentally.

\begin{table}[!tb]
\centering
\begin{threeparttable}[b]
\begin{tabular}{l@{\hskip 0.25in}l@{\hskip 0.25in}l@{\hskip 0.25in}l}
Parameter & Value & Units & Reference \\
\toprule
$R_0$ & $9 - 11$\tnote{${}^{\dagger}$} & $\text{$\mu$m}$ & \cite{Bill1972} \\
2$a$ & $0.5$\tnote{${}^{\spadesuit}$}  & $\text{$\mu$m}$ & \cite{Grierson1978} \\
$\sigma_{\mathrm{m}}$ & $40$ & $\text{pN/$\mu$m}$ & \cite{Tinevez2009} \\
$\sigma_{\mathrm{c}}$ & $374$ & $\text{pN/$\mu$m}$ & \cite{Tinevez2009} \\
$K_{\mathrm{m}}$ & $10^5$ & pN/$\mu$m & \cite{Boal2012} \\ 
$d$ & $5-12$\tnote{${}^{\clubsuit}$} & $\mu$m & \cite{shi2018} \\ 
$\epsilon^*$ & $0\% - 100\%$ & adim & \cite{Latorre2018} \\
$\tau_{\mathrm{c}}$ & $1$ & $\text{s}$ & \cite{Charras2006} \\
$\nu$ & $2.5 \cdot 10^4$ & $\text{pN} \cdot \text{s}/\mu \text{m}^2$ & \\ 
\botrule
\end{tabular}
\begin{tablenotes}
\item[${}^{\dagger}$]{Human Schlemm's canal cells are spindle-shaped with length, width and height, respectively, $100$ $\mu$m, $4-8$ $\mu$m and $5$ $\mu$m \cite{Bill1972}. 
Assuming elliptic shape, their surface area is $900-1110$ $\mu$m${}^2$. Setting it equal to $3\pi R_0^2$ yields the given estimate for $R_0$.\\}
\item[${}^{\spadesuit}$]{The estimate reported was measured in ref.~\cite{Grierson1978} at physiological intraocular pressure, 15 mmHg. 
This corresponds to a pressure drop about 7 mmHg. In general, the meshwork pore width is shown to be dependent on the imposed pressure drop.\\}
\item[${}^{\clubsuit}$]{Membrane tension was shown in ref.~\cite{shi2018} to propagate diffusively with diffusion coefficient $D=0.024$ $\mu$m${}^2$/s. 
The characteristic lifetime of GVs is of the order of a few tens of minutes, say $t \approx 10-50$ min. The estimate for the radius of the membrane patch where tension equilibrates is given by $\sqrt{2Dt}$.}
\end{tablenotes}
\caption{Summary of the model parameters.}\label{tab:parameters}\label{table2}
\end{threeparttable}
\end{table}

\section{Acknowledgements} 

A.C. benefited from a Science Research Fellowship granted by the Royal Commission for the Exhibition of 1851. A.S. was supported by the BBSRC Doctoral Training Partnership programme (Grant No. BB/M011178/1).

\section{Author contributions}
A.C., A.S., D.R.O., and C.F.L. designed research; A.C. performed research; A.C., and C.F.L. wrote the paper; A.C., A.S., D.R.O., and C.F.L. reviewed the final manuscript.


%

\end{document}